\documentclass{LMCS} 

\usepackage{amssymb}
\usepackage{amsmath}
\usepackage{exscale}
\usepackage{enumerate}
\usepackage{stmaryrd}

\usepackage{ifthen}
\usepackage{graphicx}    
\usepackage{hyperref}

\usepackage[all]{xy}
        {\begin{enumerate}
         
         }%
        {\end{enumerate}}

\newcommand{\new}[1]
        {\marginpar{\small\bfseries\ifthenelse{\equal{#1}{}}{!}{#1}}}

\renewcommand{\phi}{\varphi}
\renewcommand{\epsilon}{\varepsilon}

\DeclareMathOperator{\free}{free}

\DeclareMathOperator{\TC}{TC}
\DeclareMathOperator{\Reach}{Reach}
\DeclareMathOperator{\sph}{-sph}
\DeclareMathOperator{\dist}{dist}
\DeclareMathOperator{\ind}{index}

\def\doi{3 (4:5) 2007}
\lmcsheading%
{\doi}
{1--18}
{}
{}
{Dec.~23, 2004}
{Nov.~\phantom{0}5, 2007}
{}

\begin{document}

\title[Model Checking Synchronized Products]{Model Checking 
Synchronized Products of Infinite Transition Systems\rsuper *}

\author[S. W\"ohrle]{Stefan W\"ohrle}	
\address{Informatik 7, RWTH Aachen, 52056 Aachen, Germany}
\email{\{woehrle,thomas\}@informatik.rwth-aachen.de}  

\author[w.~Thomas]{Wolfgang Thomas}	
\address{\vskip -6 pt}	
%
%
%
%
\titlecomment{{\lsuper *}A
preliminary version of the paper appeared in 19th IEEE Symposium on Logic
in Computer Science, Turku, July 2004 \cite{wt04}.}


\begin{abstract}
  \noindent 
Formal verification using the model checking paradigm
has to deal with two aspects: The system models are structured,
often as products of components, and the specification logic has to be expressive
enough to allow the formalization of reachability properties. The present 
paper is a study on what can be achieved for infinite transition systems 
under these premises. As models we  consider products of infinite transition 
systems with different synchronization constraints. We introduce
finitely synchronized transition systems,  i.e. product systems which contain 
only finitely many (parameterized) synchronized transitions, and show that the decidability of 
FO(R), first-order logic extended by reachability predicates, of the product 
system can be reduced to the decidability of FO(R) of the components.
This result is optimal in the following sense: (1) If we allow  semifinite synchronization, i.e. just in one component infinitely many 
transitions are synchronized, the FO(R)-theory of the product system is in general 
undecidable. (2) We cannot extend the expressive power of the logic 
under consideration. Already a weak extension of first-order logic with transitive 
closure, where we restrict the transitive closure operators to arity one and 
nesting depth two, is undecidable for an asynchronous (and hence finitely synchronized) 
product, namely for the infinite grid. 
\end{abstract}

\maketitle





\section{Introduction}

In the theory of algorithmic verification, a standard framework for 
modeling systems is given by finite transition systems (often in the form 
of Kripke structures). Much effort is presently spent on extending this framework 
to cover infinite transition systems, and to deal adequately with the 
internal structure of the systems under consideration, such as their 
composition from several components. The present paper is a study on 
the scope of algorithmic model checking over transition systems that 
are composed from infinite 
components as products with various constraints on the synchronization of 
their transitions.

We consider transition graphs in the format $G=(V,(E_a)_{a \in \Sigma})$
where $V$ is the set of states (or vertices) and $E_a \subseteq V \times V$ 
the set of $a$-labeled transitions. 
The direct product of two transition graphs has an $a$-labeled transition from 
$(p,q)$ to $(p', q')$ if there are such transitions from $p$ to $p'$ 
and from $q$ to $q'$. This is the case of complete synchronization. 
The other extreme is the asynchronous product, where a transition in one 
component does not affect the other components. A main result below deals 
with the ``intermediate'' case where the component graphs are infinite and 
in each component 
only finitely many transitions are used for synchronization. We call these product
structures ``finitely synchronized''. They arise whenever the local 
computations in the components involve infinite state-spaces but 
synchronization is restricted to a finite number of actions in each 
 component. 

We study the model checking problem for products of transition graphs 
with respect to several logics that are extensions of first-order logic FO.  
A basic requirement in verification is that reachability properties
should be expressible. There are numerous ways to extend FO by features
that allow to express reachability properties. We consider here four 
extensions that cover reachability relations, listed in the order 
of increasing expressiveness:

\begin{itemize}
\item Reachability logic FO(R), which is obtained from FO-logic by 
 adjoining transitive closure operators $\Reach_\Gamma$ over subsets $\Gamma$ of 
 edge relations.
\item FO(Reg) as a generalization of FO(R) in which path labels have to match 
 a given regular expression.
\item Transitive closure logic over binary relations, which allows to proceed 
 from any definable relation (and not just from some edge relations) to 
 its transitive closure.
\item Monadic second-order logic MSO, which results from FO-logic by adjoining 
 variables and quantifiers for sets (and in which transitive closure over 
 binary relations can be expressed).
\end{itemize}

The purpose of this paper is to analyze for which types of products 
and for which of these logics $\mathcal{L}$
the decidability of the model checking problem for a  
product can be inferred from the 
decidability of the corresponding model checking problem for the components. 
In other words, we analyze for which kinds of products the decidability 
of the $\mathcal{L}$-theory of the product can be derived from the 
decidability of the $\mathcal{L}$-theories of the components. 

Our first result is such a transfer result for the logic FO(R) over 
finitely synchronized products of transition graphs. For this, we 
use a technique of ``composition'' which resembles the method of 
Feferman and Vaught \cite{fv59} in first-order model theory 
(see 
\cite{ck73}, \cite{hodg93} for introductions and \cite{Ma04} for 
a comprehensive survey). The Feferman-Vaught method (applied to FO) 
allows 
to determine the FO-theory of a product structure (e.g., 
a direct product) from the FO-theories of the components and 
some additional information on the index structure. Our proof involves 
a more detailed semantic analysis of the components, thereby exploiting the 
assumption on finite synchronization. The result 
extends a theorem of Rabinovich \cite{rabino07} on propositional modal 
logic extended by the modality EF over asynchronous products.

We show that our result is optimal in two ways.

Firstly, the result does not extend to a case where we allow a slight 
liberalization of the constraint on finite synchronization: We consider 
``semi-finite synchronization'', 
in which all components except one can synchronize via finitely many 
transitions. In the presence of a single component with infinitely many 
synchronizing transitions we may obtain a structure with undecidable 
FO(R) model checking problem, whereas the problem is decidable for the 
components individually. 
 
Secondly, we investigate whether the logic FO(R) can be extended in the 
above mentioned preservation result. For a strong extension like MSO
it is clear that decidability of the component theories does not carry over 
to the theory of the product system. As is well-known, we may work  with 
the asynchronous product of the successor structure of the natural 
numbers, which is the infinite $(\omega \times \omega)$-grid. 
(Note that the asynchronous product is finitely synchronized
with an empty set of synchronizing transitions.) The grid has an undecidable 
monadic theory, whereas the component structures have decidable monadic theories. 

We clarify the situation for weaker extensions of FO(R), 
namely FO(Reg) and transitive closure logic. We show that 
asynchronous products do not preserve the decidability of the FO(Reg)-theory. 
For transitive closure logic this undecidability result can already be obtained
for a very simple example of an asynchronous product, namely the infinite grid 
as considered above.
Moreover, we show that this undecidability phenomenon 
only appears when the TC-operator is nested. For the fragment of 
transitive closure logic with unnested TC-operators interpreted over the 
infinite grid,  we obtain a reduction to Presburger arithmetic and hence 
the decidability of the corresponding theory. 

These undecidability results complement a theorem of Rabinovich \cite{rabino07}
where the corresponding fact is shown for propositional modal logic extended by the
modality EG over finite grids.

In our results the component structures are assumed to have a decidable 
theory in one of the logics considered above. Let us summarize some of the
relevant classes and their closure properties with respect to synchronization.

A fundamental result is that pushdown graphs have
a decidable monadic second-order theory \cite{ms85}. Since then
several extensions like prefix recognizable graphs \cite{ca96} or 
Caucal graphs \cite{cau02} have been considered, see \cite{tho03} for an overview. 
These classes form an increasing sequence in this order, and
all of them enjoy a decidable MSO-theory. None of these classes is closed under 
asynchronous products. 

Two other classes of infinite graphs we like to mention are the graphs of 
ground term rewriting systems \cite{col02}  for which the FO(R)-theory is decidable, 
and ground tree rewriting systems \cite{loe02} 
for which a temporal logic with reachability and recurrence operators is decidable. 
Both classes are closed under asynchronous products.

Classes which are closed under synchronized products are rational graphs
\cite{mo00}, graphs of Thue specifications \cite{pay00}, or graphs
of linear bounded machines \cite{kp99}. However for all these
classes already the FO-theory is undecidable and hence they 
are not suitable for model checking purposes.


The paper is organized as follows. In Section 2 we give the definition
of a synchronized product of a family of graphs or transition systems, 
recall the definition of transitive closure logic, and define FO(R) and FO(Reg).

In Section 3 we show the composition theorem for finitely synchronized 
products and reachability logic and prove that this result cannot be extended 
to FO(Reg) or semifinite synchronization in general.

In Section 4 we investigate transitive closure logic over the infinite grid.
We show that if we allow transitive closure operators of arity one without 
parameters but of nesting depth two the theory of the grid is undecidable.
On the other hand we show that if no nesting of transitive closure operators is
allowed, the respective theory is decidable even in presence of parameters
in the scope of the transitive closure operators. 

\section{Preliminaries}

Let $(V_i)_{1 \le i \le n}$ be a family of sets. We denote by 
$\vartimes_{1 \le i \le n} V_i$ the Cartesian product of these sets.
Tuples $(v_1,\ldots, v_n) \in \vartimes_{1 \le i \le n} V_i$ are usually denoted
by $\bar v$, and the $i$th component of $\bar v$ as $v_i$ .

Let $\Sigma$ be a finite set of labels. A \emph{transition system} is a 
$\Sigma$-labeled directed graph $G=(V^G, (E_a^G)_{a \in \Sigma})$ where $V^G$ 
is the set of vertices of $G$ and $E_a^G \subseteq V^G \times V^G$
denotes the set of $a$-labeled edges in $G$.

\subsection{Synchronized Products}

For $ 1 \le i \le n$ let $G_i:=(V_i,(E_a^i)_{a \in \Sigma_i})$ be a 
$\Sigma_i$-labeled graph. We assume that $\Sigma_i$ is partitioned into
a set $\Sigma_i^l$ of \emph{local} labels (or actions) and a set $\Sigma_i^s$ of
\emph{synchronizing} labels, and to avoid notational complication we require 
the sets of local labels to be pairwise disjoint. An asynchronous transition 
labeled by $a \in \Sigma_i^l$ is applied only in the $i$-th component of a
state $(v_1,\ldots,v_n)$ of the product graph while the other components stay 
fixed. For synchronizing transitions we distinguish explicitly between the 
components where a joint change of states is issued and the components where the 
state does not change. To describe the latter, define
$E_\epsilon^i:=\{(v,v) \mid v \in V_i\}$ and $\tilde\Sigma_i^s:=\Sigma_i^s \cup 
\{\epsilon\}$. A \emph{synchronization constraint} is a set $C \subseteq 
\vartimes_{1 \le i \le n} \tilde\Sigma_i^s$. If $\bar c \in C$, a $\bar c$-labeled 
transtition induces a simultaneous change in the components $i$ where $c_i \neq \epsilon$
while the states do not change in the other components.

Formally, the \emph{synchronized product} of $(G_i)_{1 \le i \le n}$ defined
by $C$ is the graph $G$ with vertex set $V:=\vartimes_{1 \le i \le n} V_i$, 
asynchronous transitions with labels $a \in \bigcup_{1 \le i \le n} \Sigma_i^l$
defined by $E_a^G \bar v \bar w$ if $E_a^i v_i w_i$ and
$v_j=w_j$ for $j \neq i$, and synchronized transitions with labels $\bar c \in 
C$ defined by $E_{\bar c}^G \bar v \bar w$ if $E_{c_i}^i v_i w_i$ for every $1 \le i \le n$.
We denote the set of local transitions labels $\bigcup_{1 \le i \le n} \Sigma_i^l$ of $G$ 
by $\Sigma^l$, and the set $C \cup \Sigma^l$ of all transition labels by $\Sigma$.
A product is asynchronous if $C=\emptyset$.

Note that we slightly deviate from the definition in \cite{arn94} 
since we require the sets of local labels and synchronizing labels to be 
disjoint, and implicitly assume an asynchronous behavior of local transitions.
 
Let  $(G_i)_{1 \le i \le n}$ be a family of graphs and 
$C \subseteq \vartimes_{1 \le i \le n} \tilde\Sigma_i^s$ be a synchronization
constraint. For $\bar c \in C$ let $X_{\bar c}:=\{i \mid c_i \neq \epsilon\}$. 
For $C' \subseteq C$ we write $X_{C'} = \bigcup_{{\bar c} \in C'} X_{{\bar c}}$.
Define 
\[\bar u \sim_{\bar c} \bar v:\Leftrightarrow u[X_{\bar c}]=v[X_{\bar c}],\]
i.e. $\bar u \sim_{\bar c} \bar v$ if $\bar u$ and $\bar v$ agree on the synchronizing components.
The synchronized product $G$ of $(G_i)_{1 \le i \le n}$ defined by $C$ is called 
\emph{finitely synchronized} if $\ind(\sim_{\bar c})$, i.e. the number 
of equivalence classes of $\sim_{\bar c}$, is finite for every $\bar c \in C$.
In the conference version \cite{wt04} of this paper, finitely synchronized products involve only finitely 
many individual synchronizing transitions, thus disallowing the label $\epsilon$ in
the synchronization constraint. In the present treatment we allow finitely many parametrized 
synchronized transitions: The inclusion of constraints $\bar c$ with $c_i=\epsilon$ means that 
in the $i$-th component the transition $\bar c$ applies to arbitrary states of $V_i$ and hence 
possibly infinitely many individual synchronizing transitions may be present in a finitely 
synchronized product\footnote{Thus, the proof of Theorem \ref{theo:compos} below involves more technicalities
than the corresponding proof in \cite{wt04}.}.

We collect some technical preparations in the subsequent Lemma \ref{lem:equiv-classes}. For this
we define for every $\emptyset \neq C' \subseteq C$ the eqivalence relation 
\[\bar u \sim_{\bar C'} \bar v:\Leftrightarrow \bar u \sim_{\bar c} \bar v \text{ for every } \bar c \in C'\]
and restrict the relation $\sim_{C'}$ to the set of vertices of the synchronized product from
which an outgoing transition exists for every $\bar c \in C'$, i.e. to the set
\[V_{C'}:=\{\bar u \in \vartimes_{1 \le i \le n} V_i \mid \forall \bar c \in C' \ \exists \bar v \text{ such that } 
(\bar u,\bar v) \in E_{\bar c}\}.\]

\begin{lem}\label{lem:equiv-classes}
Let  $(G_i)_{1 \le i \le n}$ be a family of graphs and 
$C \subseteq \vartimes_{1 \le i \le n} \tilde\Sigma_i^s$ be a synchronization
constraint. 
\begin{enumerate}[(a)]
\item If $G$ is finitely synchronized, then $\ind(\sim_{C'})$ is finite for every $\emptyset \neq C' \subseteq C$.
\item For every subset $\emptyset \neq C' \subseteq C$,
if $u \sim_{C'} v$ and $G \models \Reach_{\Sigma^l \cup C'}[u,w]$
there exists a $w'$ such that $G \models \Reach_{\Sigma^l \cup C'}[v,w']$
and $w \sim_{C'}  w'$.
\item Let $\Gamma \subseteq \Sigma^l \cup C'$. If 
 $G \models \Reach_\Gamma[u,v]$, $G \models \Reach_\Gamma[v,w]$
 and $u \sim_{C'} v$ then $G \models \Reach_\Gamma[u, w]$ and the path from 
 $u$ to $w$ can be chosen such that no intermediate vertex is $\sim_{C'}$-equivalent to 
 $u$.
\end{enumerate}
\end{lem}

\proof
(a) If $G$ is finitely synchronized, then $\ind(\sim_{\bar c})$ is finite for every $\bar c \in C$.
    If $C' \subseteq C'' \subseteq C$ then $C''$ refines $C'$ on $V_{C''} \subseteq V_{C'}$. Therefore, 
    for every $C' \subseteq C$ the number of equivalence classes of $\sim_{C'}$ is bounded by
    $\vartimes_{\bar c \in C}\ind(\sim_{\bar c})$.

(c) is a direct consequence of (b) which remains to be 
    shown. Let $u \sim_{C'} v$ and $G \models \Reach_{\Gamma}[u,w]$.
    Since transitions labeled with  symbols from $\bigcup_{i \notin X_{C'}} \Sigma_i^l$ 
    commute with transitions labeled by symbols from  $\bigcup_{i \in X_{C'}} \Sigma_i^l \cup C'$
    we may w.l.o.g. assume that the path from $u$ to $w$ is of the form
    \[u=u_1 \xrightarrow{a_1} u_1 \xrightarrow{a_2} \ldots \xrightarrow{a_{m-1}} u_m = u_1' \xrightarrow{b_1} u_2'
      \xrightarrow{b_2} \ldots \xrightarrow{b_{n-1}} u_n' = w\]
    and $a_j \in \bigcup_{i \notin X_{C'}} \Sigma_i^l$ for $1 \le j \le m$ and $b_j \in 
    \bigcup_{i \in X_{C'}} \Sigma_i^l \cup C'$ for $ 1 \le j \le n$.
    Hence by definition of $\sim_{C'}$ we have $u[X_{C'}]=u_1'[X_{C'}]=v[X_{C'}]$. Thus
    there is a path $v'=v_1 \xrightarrow{b_1} v_2  \xrightarrow{b_2} \ldots \xrightarrow{n-1} v_n =w'$ in $G$
    and $w \sim_{C'} w'$.
\qed

\subsection{First-Order Logic and Extensions}

We assume that the reader is familiar with first-order logic FO 
over graphs. We denote formulas by $\phi(x_1,\ldots,x_n)$ to express that 
the free variables of $\phi$ are among $x_1,\ldots,x_n$. If $G$ is a graph and
$v_1,\ldots,v_n$ are the vertices assigned to the variables $x_1,\ldots,x_n$, we
denote by $(G,v_1,\ldots,v_n) \models \phi(x_1,\ldots,x_n)$ or shortly by
$G \models \phi[v_1,\ldots,v_n]$ that the formula $\phi$ is satisfied in $G$
under the respective variable assignment.

\emph{Transitive closure logic} FO(TC) is defined by extending 
FO with formulas of the type 
\[\psi:=\left[\TC_{\bar x, \bar y} \phi(\bar x, \bar y, \bar z)\right]\bar s, \bar t\]
where $\phi(\bar x, \bar y, \bar z)$ is a FO(TC)-formula, $\bar x, \bar y$ are disjoint 
tuples of free variables of the same length $k>0$,  $\bar s, \bar t$ are tuples of 
variables of length $k$ and $\free(\psi):=(\free(\phi)\setminus \{\bar x, \bar y\})
\cup \{\bar s, \bar t\}$. Note that in the notation 
$\left[\TC_{\bar x, \bar y} \phi(\bar x, \bar y, \bar z)\right]\bar x, \bar y$
the variables inside the square brackets are bound while the variables at the end of the formula
occur free.

Let $G$ be a graph, let $\bar c$, $\bar d$, and $\bar e$ be the interpretations of the variables 
$\bar z$, $\bar s$, and $\bar t$ in $\phi$. Let $E$ be the relation on $k$-tuples defined by 
$E(\bar c):=\{(\bar a, \bar b) \mid (G, \bar a, \bar b, \bar c) \models \phi(\bar x, \bar y, \bar z)\}$,
and $E'(\bar c)$ be its transitive closure, i.e. $(\bar a, \bar b) \in E'(\bar c)$
iff there exists a sequence $\bar f_0, \bar f_1, \ldots, \bar f_l$ such that
$\bar f_0 = \bar a$, $(\bar f_i, \bar f_{i+1}) \in E(\bar c)$ for $ 1 \le i < l$, and
$\bar f_l = \bar b$. The semantics of the FO(TC)-formula above is defined by
\[(G, \bar c, \bar d, \bar e) \models \left[\TC_{\bar x, \bar y} \phi(\bar x, \bar y, 
\bar z)\right]\bar s, \bar t \Leftrightarrow  (\bar d, \bar e) \in E'(\bar c).\]

We call the variables $\bar z$ parameters for the transitive closure operator.
By $\textrm{FO(TC)}_{(k)}$ be denote the fragment of FO(TC) where the 
transitive closure operation is only allowed to define relations over tuples of 
length $\le k$, i.e. the length of the tuples $\bar x, \bar y$ in the
definition above is bounded by $k$. For example, in $\textrm{FO(TC)}_{(1)}$
we can only define binary relations using a transitive closure operator.
For finite models the arity hierarchy $(\textrm{FO(TC)}_{(k)})_{k \ge 0}$ is 
strict \cite{gro96}.

By $\textrm{FO(TC)}_{(k)}^l$ we denote the 
fragment of $\textrm{FO(TC)}_{(k)}$ where the nesting depth of transitive closure 
operations is bounded by $l$. 

In transitive closure logic we can express that from a vertex
$x$ a vertex $y$ is reachable via a path with labels from some set 
$\Sigma' \subseteq \Sigma$ by 
\[\Reach_{\Sigma'}(x,y):=\Big[\TC_{x,y} \big(x=y \vee \bigvee_{a \in \Sigma'} E_a xy\big)\Big]x,y.\] 

We call the restriction of FO(TC) where the only transitive closure formulas allowed are
of the form $\Reach_{\Sigma'}(x,y)$ for  $\Sigma' \subseteq \Sigma$ \emph{reachability logic} 
and denote it by FO(R).

 The expressive power of the reachability predicates in FO(R) is limited, 
 e.g. we cannot express that there is a path between vertex $v$ and $w$ in 
 the graph whose labels form a word in a given regular language. 

 We denote by FO(Reg) first-oder logic extended by reachability predicates 
 $\Reach_r(x,y)$ for regular expressions $r$ over $\Sigma$, where 
 $G \models \Reach_r[v,w]$ if there is a path in $G$ from $v$ to $w$
 labeled by a word contained in the language described by $r$.



\section{Synchronization and FO(R)}

In this section we show that synchronization preserves the decidability 
of the FO(R)-theory if (and only if) the product is finitely synchronized.
For this case we prove a composition theorem that reduces the evaluation
of a formula in the product graph to the evaluation of several formulas in
the component graphs and a Boolean combination of these truth values.
This result does not extend to the case of FO(Reg).

Furthermore we show that \emph{semifinite} synchronization of two components,
where in just  one of the components infinitely many edges are allowed to
be synchronized, does in general not preserve the decidability of the FO(R)-theory.

\begin{thm}\label{theo:compos}
Let $G$ be a finitely synchronized product of a family $(G_i)_{1 \le i \le n}$ of graphs 
with decidable FO(R)-theories. Then the FO(R)-theory of $G$ is also decidable, and for an
FO(R)-formula $\phi$ we can effectively construct sets of formulas $\Psi_i$ and a Boolean 
formula $\alpha$ such that $G \models \phi$ iff $\alpha$ is true under an Boolean interpretation
defined by the truth values of the formulas in $\Psi_i$.
\end{thm}

\proof
Let $(G_i)_{1 \le i \le n}$ be a family of graphs whose signatures 
$\Sigma_i := \Sigma_i^l \cup \Sigma_i^s$ are partitioned into local and synchronizing labels.  
Let $C \subseteq \vartimes_{1 \le i \le n} \tilde\Sigma_i^s$ be a synchronization constraint
such that the product $G$ of $(G_i)_{1 \le i \le n}$ is finitely synchronized with respect
to $C$.

We show by induction that for every FO(R)-formula over $\Sigma$ there are finite sets 
$\Psi_i$ of $\Sigma_i$-formulas and a Boolean 
formula $\alpha$ over predicates $p_i(\psi_j^i)$ $(1 \le i \le n,\ 1 \le j \le |\Psi_i|)$
such that 
\begin{equation}\label{eq:theo1}
(G, \bar v_1,\ldots \bar v_m) \models \phi(x_1,\ldots, x_m) \Leftrightarrow
  I(\bar v_1, \ldots, \bar v_m) \models \alpha
\end{equation}
where $I(\bar v_1, \ldots, \bar v_m)$ is the Boolean interpretation defined by 
\[I(\bar v_1, \ldots, \bar v_m)(p_i(\psi_j))= 
  \begin{cases} 
  \text{true} & \text{if } (G_i,v_1^i,\ldots,v_m^i) \models \psi_j^i \\
  \text{false} & \text{otherwise.}
  \end{cases}
\]
We start with the atomic formulas. For $x=y$ let $\psi_i:=(x=y)$, for $E_a xy$ 
with $a \in \Sigma_i^l$ let $\psi_i:=E_axy$ and $\psi_j:=(x=y)$ for $i \neq j$, and
for $E_{\bar c} xy$ with $\bar c \in C$ let $\psi_i:=E_{c_i}xy$. 
For every formula above let $\alpha:=\bigwedge_{1 \le 
i \le n} p_i(\psi_i)$. Obviously (\ref{eq:theo1}) holds in all cases, so the remaining
``atomic'' formulas we have to take care of are of the form $\Reach_{\Gamma}(x,y)$
for $\Gamma \subseteq \Sigma$. 

For this part of the proof we proceed by induction on the number of synchronizing transitions from 
$C$ which appear in $\Gamma$. We may assume that $\Gamma$ comprises all local transition labels,
i.e. that $\Sigma^l \subseteq \Gamma$; otherwise in the following every occurrence of $\Sigma_i^l$ has to be 
replaced by $\Sigma_i^l \cup \Gamma$.

We first consider the case that there is only a single synchronizing 
transition $\bar c \in \Gamma$. By the definition of finitely synchronized 
product 
we know that $\ind(\sim_{\bar c})$ is 
finite, and by Lemma   \ref{lem:equiv-classes} (c) that we have to pass through every equivalence class
at most once. Let $k = \ind(\sim_{\bar c})$. For $i \in X_{\bar c}$ and $1 \le m \le k$ define
\begin{align*}
\psi^i_{(\bar c, m)}(x,y):= \exists z_1 \ldots \exists z_m \Big( & \Reach_{\Sigma_i^l}(x,z_1) \wedge
  y=z_m \\ 
 & \wedge \bigwedge_{1 \le i < m} \exists w \big( E_{c_i} z_i w \wedge \Reach_{\Sigma_i^l}(w, z_{i+1}) \big)\Big)
\end{align*}
which expresses that on a path from $x$ to $y$ in component $i$ exactly $m$ vertices $z_1, \ldots, z_m$
are passed from which a synchronized transition is possible.
For $i \notin X_{\bar c}$ we set 
\[\psi^i_{\bar c, m}(x,y):= \Reach_{\Sigma_i^l}(x,y)\]
and define $\Psi_i(\bar c):=\{\psi^i_{(\bar c, m)}(x,y) \mid 1 \le m \le k\}$. 
Setting 
\[\alpha(\bar c):=\bigvee_{1 \le m \le k} \bigwedge_{1 \le i \le n} p(\psi^i_{\bar c, m})\] 
ensures (\ref{eq:theo1}) for sets $\Gamma$ which contain at most one synchronizing edge label $\bar c$.

Let now $C' = C \cap \Gamma$. By the induction hypothesis we may assume that for every subset $C'' \subset C'$ 
there are families of formulas $\Psi_i(C''):=\{\psi_{(C'',m)}^i(x,y) \mid m \le m(C'')\}$ and Boolean formulas 
$\alpha(C'')$ such that (\ref{eq:theo1}) holds, i.e.
\begin{equation}\label{eq:prooftheo1}
  G \models \Reach_{C'' \cup \Sigma^l}[\bar v,\bar w ] \Leftrightarrow
  I(\bar v, \bar w) \models \alpha(C'').
\end{equation}
Let $k:=\ind(\sim_{C'})$ and $l:=\sum_{C'' \subset C'} \ind(\sim_{C''})$, 
for $1 \le r \le k$ let $\sigma_1$ be a mapping $\sigma_1: \{1,\ldots, r\} \rightarrow \{1, \ldots, l\}$
and $\sigma_2$ a mapping $\sigma_2: \{1,\ldots,l\} \rightarrow \{(C'',s) \mid C'' \subset C',\ s \le m(C'')\}$.
The number of vertices in $V_{C'}$ which are passed on the path from vertex $\bar u$ to $\bar w$ is $r$. The 
mapping $\sigma_1$ then determines the number of $\sim_{C''}$ equivalence classes which are passed on the path 
between consecutive vertices in $V_{C'}$ and $\sigma_2$ determines the order in which vertices from  $\sim_{C''}$
eqivalence classes appear.

Let $\pi_1, \ldots, \pi_t $ be an enumeration of all mappings which can be obtained by composing the mappings 
$\sigma_1$ and $ \sigma_2$. We define for $ t' \le t$, $\pi_{t'}:= \sigma_2 \circ \sigma_1$ with $\sigma_j$ as above 
and $1 \le i \le n$ the formula
\[
   \begin{array}{ll}
   \psi^i_{(C',t')}(x,y):=  \exists y_1 \ldots y_r & \Big[ y_1 = x \wedge y_r = y \\ 
    &\wedge \bigwedge_{1 \le p < r} \Big(\exists z_1 \ldots z_{\sigma_1(p)} \big( z_1 =y_p \wedge z_{\sigma_1(p)}=y_{p+1} \\ 
    & \wedge \bigwedge_{1 \le q < \sigma_1(p)} \psi^i_{\sigma_2(q)}(z_q,z_{q+1}) \big) \Big) \Big].
   \end{array}
\]
The Boolean formula $\alpha(C')$ is then defined to be 
\[\alpha(C'):= \bigvee_{1 \le t' \le t} \bigwedge_{1 \le i \le n} p(\psi^i_{(C',t')}).\]

We claim now that for every $C' \subseteq C$
\begin{equation}\label{eq:comp-to-show}
 G \models \Reach_{\Sigma^l \cup C'}[\bar u, \bar v] \Leftrightarrow I(\bar u,\bar v) \models \alpha(C').
\end{equation}

We first consider the direction from right to left. Let $I(\bar u,\bar v) \models \alpha(C')$.
The case $C'=\{\bar c\}$ has already been dealt with above. So assume that (\ref{eq:comp-to-show}) holds 
for every $C'' \subset C'$. Then $I(\bar u,\bar v) \models \bigwedge_{1 \le i \le n} p(\psi^i_{C',t'})$ for
some $t'$, i.e there exits an $r$ and mappings $\sigma_1: \{1, \ldots, r\} \rightarrow \{1, \ldots, l\}$
and $\sigma_2:  \{1, \ldots, l\} \rightarrow \{(C'',s) \mid C'' \subset C',\ s \le m(C'')\}$ such that
for $1 \le i \le n$
\[ 
\begin{array}{ll}
 (G,u_i,v_i) \models \exists y_1 \ldots y_r & \Big[ y_1 = x \wedge y_r = y \\ 
    & \wedge \bigwedge_{1 \le p < r} \Big( \exists z_1 \ldots z_{\sigma_1(p)} \big( z_1 =y_p \wedge z_{\sigma_1(p)}=y_{p+1} \\ 
    & \wedge \bigwedge_{1 \le q < \sigma_1(p)} \psi^i_{\sigma_2(q)}(z_q,z_{q+1}) \big) \Big) \Big].
\end{array}
\]
If we denote the the valuation of the variables $z_j$ (respectively $y_j$) in $G_i$ which make the formula above true by $z_j^i$ 
(respectively $y_j^i$) and their $n$-tuple by $\bar z_j$ (respectively $\bar y_j$) we obtain that 
$I(\bar z_j,\bar z_{j+1}) \models \alpha(\sigma_2(q)_1)$ for $1 \le j < \sigma_1(p)$ (here $\sigma_2(q)_1$ denotes the
first component of $\sigma_2(q)$). Hence
$G \models \Reach_{\Sigma^l \cup \sigma_2(q)_1}[\bar z_j, \bar z_{j+1}]$ for $1 \le j \le \sigma_1(p)$ and
since $\bigcup_{1 \le q \le \sigma_1(p)} \sigma_2(q)_1 \subseteq C'$
also $G \models \Reach_{\Sigma^l \cup C'}[\bar y_j, \bar y_{j+1}]$ for $1 \le j \le r$.
Hence we obtain $G \models \Reach_{\Sigma^l \cup C'}[\bar u, \bar v]$.

For the direction from left to right suppose that $G \models \Reach_{\Sigma^l \cup C'}[\bar u, \bar v].$
By Lemma \ref{lem:equiv-classes} (c) we know that there is a path from $\bar u$ to $\bar v$ in $G$ which
passes every $\sim_{C'}$ equivalence class ot most once. Let $\bar y_1, \ldots, \bar y_r$ be the
sequence of these vertices from $V_{C'}$ on the path. We now consider for $1 \le j < r$ the path segments
between $\bar y_j$ and $\bar y_{j+1}$. Every such path segment can be further decomposed in  the following 
way: Let $\bar z_1$ be the first vertex in the segment which is contained in some $V_{C''}$ for $\emptyset \neq C'' 
\subset C'$. If there is no such $\bar z_1$ only local labels can appear on the path from $\bar y_j$ to 
$\bar y_{j+1}$. In this case choose $\bar z_1:= \bar y_{j+1}$. 

Then we choose $\bar z_2$ to be the last vertex
on the path from $\bar y_j$ to $\bar y_{j+1}$ such that $G \models \Reach_{\Sigma^l \cup C''}[\bar z_1, \bar z_2]$,
i.e. $z_2 \in V_{C'''}$ for some $C''' \subset C'$ with $C''' \setminus C'' \neq \emptyset$.
This decomposition can be continued until $\bar y_{j+1}$ is reached. 

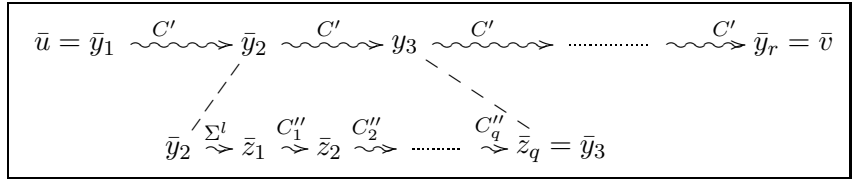
\begin{figure}[htbp]
  \centering
  \fbox{
  \xymatrix@C=1pc{
   \bar u= \bar y_1 \ar@{~>}[rr]^{C'} && \bar y_2 \ar@{~>}[rr]^{C'} \ar@{--}[dl]
     && y_3 \ar@{~>}[rr]^{C'} \ar@{--}[drr] && \ar@{..}[r] & \ar@{~>}[rr]^{C'} && \bar y_r=\bar v \\
   & \bar y_2 \ar@{~>}[r]^{\Sigma^l} & \bar z_1 \ar@{~>}[r]^{C_1''}& \bar z_2 \ar@{~>}[r]^{C_2''}  
   & \ar@{..}[r] & \ar@{~>}[r]^>>>{C''_q} & \bar z_q = \bar y_3  }
  }
  \caption{Sample decomposition of a path}
  \label{fig:path-decomp}
\end{figure}

Figure \ref{fig:path-decomp} shows such a decomposition of a path from $\bar u$ to $\bar v$.
Every path segment from $\bar y_j$ to $\bar y_{j+1}$ is again partintioned as shown. For
sake of readability we mention only the set of synchronizing labels allowed on the intermediate 
paths and write $C'$ for $C' \cup \Sigma^l$.

By Lemma \ref{lem:equiv-classes} (c)
we again know that the number of intermediate vertices $\bar z$ can be bounded by 
$l:=\sum_{C'' \subset C'} \ind(\sim_{C''})$. By the induction hypothesis on subsets $C'' \subset C'$ 
we know that for every pair of successive vertices $\bar z_j, \bar z_{j+1}$ with $\bar z_j \in V_{C''}$
there exists a conjunct of $\alpha(C'')$, i.e. some $s$ such that
\[ G \models \Reach_{C'' \cup \Sigma^l}[\bar z_j, \bar z_{j+1}] \Rightarrow 
   I(\bar z_j, \bar z_{j+1}) \models \bigwedge_{1 \le i \le n} p(\psi^i_{(C'',s)}).\] 
In particular we have $G_i \models \psi^i_{(C'',s)}[z_j^i,z_{j+1}^i]$ for every $1 \le i \le n$
and all inermediate vertices $z_j$. 

Combining these decomposition results we obtain that there exists some $r$ bounded by $\ind(\sim_{C'})$ (the 
number of vertices $\bar y$), a function $\sigma_1:\{1,\ldots,r\} \rightarrow \{1,\ldots, l\}$ 
which determines the number of intermediate vertices $\bar z$ between the $\bar y$ vertices, and a function 
$\sigma_2:\{1,\ldots,l\} \rightarrow \{(C'',s) \mid C'' \subseteq C'', \ s \le m(C'')\}$ which determines
to which $V_{C''}$ an intermediate vertex $\bar z_j$ belongs and which conjunct of $\alpha(C'')$ is satisfied
by the interpretation induced by $\bar z_j$ and $\bar z_{j+1}$.  Thus we obtain that 
$G_i \models \psi^i_{(C',s)}[u_i,v_i]$ for $1 \le i \le n$ and some $s$ an hence
$I(\bar u, \bar v) \models \alpha(C')$.

The finishes the proof for atomic formulas. Formulas composed by Boolean connectives and
existential quantification are now easy to handle.

The case of Boolean connectives may be solved in the standard way.
Let $\phi_1(\bar x)$ and $\phi_2(\bar y)$ be FO(R)-formulas and 
$\alpha_1$, $(\Psi_i^1)_{1 \le i \le n}$ as well as $\alpha_2$, $(\Psi_i^2)_{1 \le i \le n}$
be given by the induction hypothesis.  Then, for $\neg \phi_1(\bar x)$ we can choose the
same $(\Psi_i^1)_{1 \le i \le n}$ and the Boolean formula to be $\neg \alpha_1$, and for
$\phi_1(\bar x) \vee \phi_2(\bar y)$ we choose $\Psi_i:=\Psi_i^1 \cup \Psi_i^2$ and 
$\alpha = \alpha_1 \vee \alpha_2$.

To finish the proof let $\phi(x_1,\ldots,x_n):= \exists x_{n+1} \phi_1(x_1,\ldots,x_{n+1})$. Let 
$\Psi_i^1$ and $\alpha_1$ be the formulas computed for $\phi_1(x_1,\ldots,x_{n+1})$. Let 
$\mathcal I$ be the set of all satisfying assignments for $\alpha_1$. For every 
$I \in \mathcal I$ let $I_i:=\{j \mid I(p_i^j)=\text{true}\}$. Then sets $\Psi_i$ for $1 \le i \le n$
are constructed by adding for every $I \in \mathcal I$ the formula 
\[\psi_i^I(x_1,\ldots,x_n):= \exists x_{n+1} \Big( \bigwedge_{j \in I_i} \psi_i^j \wedge
  \bigwedge_{j \notin I_i} \neg \psi_i^j\Big).\]
Then we can define $\alpha:=\bigvee_{I \in \mathcal I} \bigwedge_{1 \le i \le n} p(\psi_i^I)$. 
\qed

For a complexity analysis of this algorithm, note that even in the special case in which the synchronization 
constraint does not contain $\epsilon$, the number of formulas which have to be evaluated in the components 
cannot be bounded by an elementary function. This is due to the exponential increase of the sets $\Psi_i$ which result
from dealing with existential quantifiers.

It is easy to see that Theorem \ref{theo:compos} also covers
FO(Reg)-formulas with regular expressions built from 
$\Gamma_i^*$ for $\Gamma_i \subseteq \Sigma$ using $\cdot$ and $+$.
However, if we allow reachability predicates with regular expressions of the 
form $(\Gamma_1 \cdot \Gamma_2)^*$ the decidability of the corresponding 
theory will be lost.


\begin{thm}
  Asynchronous products do not preserve the decidability of the 
 FO(Reg)-theory.
\end{thm}

\proof
We use a 2-PDA $\mathcal  A$ (pushdown automaton 
with two stacks) that simulates 
a universal Turing machine (cf. \cite{hu79}). 
Formally a 2-PDA is a tuple 
$\mathcal A=(Q, \Sigma, \Gamma, q_0, \Delta, f)$
where $Q$ is a finite set of states, $\Sigma$ and $\Gamma$ the input alphabet,
respectively  stack alphabet, 
$q_0$ is the initial state, $f$ is the final state, and 
$\Delta \subseteq Q \times \Sigma \times (\Gamma \cup \{\epsilon\})^2 \times (\Gamma \cup \{\epsilon\})^2 \times Q$ the transition relation. The configuration with 
state $p$ and stack contents $u,v$ (discarding the stack bottom symbols) 
is denoted by $(p,u,v)$ (similarly a pair
$(p,u)$ is a configuration of a standard PDA).  
We assume that Turing 
machines (as well as 2-PDA's) are normalized, i.e. that each state 
is reachable from the 
initial state $q_0$, the only sink state is the final state $f$ and there
are no incoming transitions to $q_0$. 

Input words for the universal 2-PDA $\mathcal A$ are  
of the form $w_1\$w_2\#$ where $w_1$ is the code 
of a Turing machine and $w_2$ an input word for the Turing machine. We 
assume that $\mathcal A$ processes such an input word in two phases: 
First $w_1\$w_2\#$ is written into the first stack 
(in reverse order) and then transferred into the second stack (with the 
first letter of $w_1$ on top of the stack). With this configuration 
the second phase starts (and we call its initial state $q_0^2$), 
realizing the actual simulation of the universal 
Turing machine. It is well-known that the reachability problem for 
$\mathcal A$ (``Given $w_1\$w_2\#$ as input, 
does $\mathcal A$ reach the final state?") is undecidable. 

To reduce this reachability problem for $\mathcal A$ to the 
model checking problem for FO(Reg) 
over an asynchronous product of graphs with decidable FO(Reg)-theory, 
 we split $\mathcal A$ into two component pushdown automata 
\begin{align*}
 \mathcal A_1 &= (Q, \Sigma \times \Delta, \Gamma, q_0, \Delta_1, f) \\
 \mathcal A_2 &= (Q, \bar \Sigma \times \bar \Delta, \Gamma, q_0, \Delta_2, \bar f) 
\end{align*}
where for every $\delta=(q,a,\gamma_1,\gamma_2,\gamma_3,\gamma_4,p) \in \Delta$
the following transitions are included:
\begin{align*}
 (q,(a,\delta),\gamma_1, \gamma_3, p) & \text{ to } \Delta_1, \\
 (q,(\bar a,\bar \delta),\gamma_2, \gamma_4, p) & \text{ to } \Delta_2.
\end{align*}

Each of the graphs generated by $\mathcal A_1$ and $\mathcal A_2$ has a decidable 
MSO-theory 
and therefore also a decidable FO(Reg)-theory. 
Let $\mathcal B$ their asynchronous product. 

Let $r$ be the regular expression 
\begin{align*}
r=\Big(\bigvee_{\substack{\delta \in \Delta \\ a \in \Sigma}} (a,\delta)(\bar a,\bar \delta)\Big)^*
\end{align*}
which states that a transition of $\mathcal A_1$ is followed by the corresponding transition of $\mathcal A_2$.

We obtain that
\[\mathcal B \models \Reach_r(x,y)[((q,u),(q,v)),((q',u'),(q'',v'))]\]
iff $q'=q''$ and $\mathcal A$ can reach from configuration $(q,u,v)$ the configuration $(q',u',v')$.

It is now easy to construct for every word $w_1\$w_2\#$ a first-order formula
$\phi_{w_1\$w_2\#}(x,y)$ such that 
\[\mathcal B \models \phi_{w_1\$w_2\#}(x,y)[((q_0,\epsilon),(q_0,\epsilon)),((q,u_1),(q,u_2))]\]
iff $u_1 = \epsilon$, $u_2 = w_1\$w_2\#$ and $q=q_0^2$.
Then we obtain that
\begin{multline*}
 \mathcal B \models \exists z_1 \exists z_2 \exists z_3 \Big( \phi_{w_1\$w_2\#}(((q_0,\epsilon),(q_0,\epsilon)),z_1)
 \wedge \Reach_r(z_1,z_2) \\ \wedge \bigvee_{\substack{\delta \in \Delta \\ a \in \Sigma}} \big( E_{(a,\delta)} z_2 z_3 
 \wedge  E_{(\bar a,\bar \delta)} z_3 ((f,u),(f,v)) \big) \Big)
\end{multline*}
iff $\mathcal A$ reaches a halting configuration after processing $w_1\$w_2\#$.
Note that since $\mathcal A$ is normalized we can ensure that the initial configuration 
$((q_0,\epsilon),(q_0,\epsilon))$
and all final configurations $((f,u),(f,v))$ are first-order definable.
%
%
\qed
We now turn to the proof that semifinite synchronization in general does not
preserve the decidability of the FO(R)-theory. We reduce the halting problem of 
deterministic Turing machines to the model checking problem for FO(R) for 
synchronized products of finite graphs and infinite graphs which are generated by  
ground tree rewriting systems (GTRS). The GTRS graphs we will construct are of finite out-degree 
and hence have a decidable FO(R)-theory \cite{loe02,loe03}.

The GTRS graph will encode computations of the Turing machine $M$, but not all
of them are valid. We will use the synchronization with a finite graph to eliminate 
computations which are not valid.

Our construction of the GTRS graph encoding computations of $M$ follows
ideas of \cite{loe03}. Before we start the proof we 
give a short definition of the Turing machine model we use and of ground 
tree rewriting systems. For a more detailed description we refer to  
\cite{hu79} and \cite{loe03}.

A \emph{deterministic Turing machine} is a tuple $M=(Q,\Gamma,q_0,q_f,\delta)$ where
$Q$ is a finite set of states, $\Gamma$ is an alphabet containing a
designated blank symbol \textvisiblespace~, $q_0$ is the initial state, $q_f$
is the halting state, and $\delta:Q \times \Gamma \rightarrow Q \times \Gamma \times
\{L,R\}$ is the transition function. A \emph{configuration} of $M$ is a sequence
$a_1, \ldots a_k, q, b_l, b_{l-1} \ldots b_1$ where $a_i,b_i \in \Gamma$, $q \in Q$
and $b_l$ denotes the symbol currently read by the head of the machine. We consider
two configurations to be equivalent if they differ only in heading or trailing blank 
symbols, and do not distinguish between equivalent configurations.

A \emph{ground tree rewriting system} is a tuple $\mathcal R=(A, \Sigma, R, t_0)$
where $A$ is a ranked alphabet, $\Sigma$ is a set of labels for the 
rules, $R$ is a finite set of rules, and $t_0$ is a finite tree over $A$.
We denote the set of all finite trees over $A$ by $T_A$. A \emph{rewriting rule}
$r$ is of the form $t \xrightarrow{b} t'$ with $t,t' \in T_A$ and $b \in \Sigma$.
A rule $r$ is applicable to a tree $s$ if there is a subtree $s_1$ of $s$ equal to $t$,
and the result of an application of $r$ to $s$ is a tree $s'$ obtained from $s$
by replacing $s_1$ with $t'$. $\mathcal R$ generates a $\Sigma$-labeled graph 
whose vertices are the trees that can be obtained from $t_0$ by applying 
rewriting rules from $R$, with a $b$-labeled edge between $s$ and $s'$ if
$s'$ results from $s$ by an application of a rule of the form  $t \xrightarrow{b} t' \in R$.

\begin{thm}\label{theo:semi-undec}
Semifinite synchronization does not preserve the decidability of 
the FO(R)-theory.
\end{thm}

\proof
Let $M=(Q,\Gamma,q_0,q_f,\delta)$ be a deterministic Turing machine.
We assume that $q_0 \neq q_f$, $Q \cap \Gamma =\emptyset$, $X \notin Q \cup \Gamma$  and
encode a configuration $a_1,\ldots, a_k, q, b_l, b_{l-1}, \ldots b_1$ of $M$ 
by a tree 
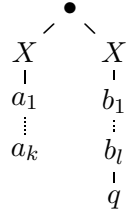
\[\xymatrix@R=0.3pc@C=0.2pc{ 
  & \bullet \ar@{-}[dl] \ar@{-}[dr] & \\
  X \ar@{-}[d] && X\ar@{-}[d] \\
  a_1 \ar@{..}[d] && b_1 \ar@{..}[d] \\ 
  a_k && b_l \ar@{-}[d] \\
  && q}
\]
Every transition of the Turing machine will be simulated by the
rewriting system in two steps, by first rewriting the right branch of
the configuration tree, and then rewriting the left branch.
The labels of the rewriting rules will indicate which letter from 
$\Gamma$ has to be added $(+)$ or removed $(-)$ from the left branch of
the configuration tree, and $\top$ respectively $\bot$ indicate
whether the halting state has been reached or not.

More precisely we define a GTRS $\mathcal R=(A,\Sigma,R,t_0)$
where $A_2=\{\bullet\}$, $A_1=\Gamma \cup \{X\}$, $A_0= A_1 \cup Q$,
$\Sigma = \{+,-\} \times (\Gamma \cup \bar \Gamma) \times\{\bot, \top\}$ and
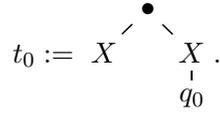
\[t_0:=\vcenter{\xymatrix@R=0.3pc@C=0.2pc{ & \bullet \ar@{-}[dl] \ar@{-}[dr] & \\ X  && X \ar@{-}[d] \\ && q_0}}.\]
The set $R$ is defined by adding for $\delta(q,b)=(p,c,L)$ and 
every $a \in \Gamma$ the rules

\[\vcenter{\xymatrix@R=0.3pc@C=0.2pc{b \ar@{-}[d] \\ q}} \xrightarrow{(-,a,*)}
  \vcenter{\xymatrix@R=0.3pc@C=0.2pc{c \ar@{-}[d] \\ a \ar@{-}[d] \\ p}}
 \text{ and } 
 \vcenter{\xymatrix@R=0.3pc@C=0.2pc{X \ar@{-}[d] \\ q}} \xrightarrow{(-,a,*)} 
 \vcenter{ \xymatrix@R=0.3pc@C=0.2pc{c \ar@{-}[d] \\ a \ar@{-}[d] \\ p}}
 \text{ if } b= \mbox{\textvisiblespace},
\]
and for $\delta(q,b)=(p,c,R)$  and every $a \in \Gamma$ the rules
\[\vcenter{\xymatrix@R=0.3pc@C=0.2pc{a \ar@{-}[d] \\  b \ar@{-}[d] \\ q}} \xrightarrow{(+,c,*)}
  \vcenter{\xymatrix@R=0.3pc@C=0.2pc{a \ar@{-}[d] \\ p}}
 \text{ and } 
 \vcenter{\xymatrix@R=0.3pc@C=0.2pc{X \ar@{-}[d] \\ q}} \xrightarrow{(+,c,*)} 
 \vcenter{ \xymatrix@R=0.3pc@C=0.2pc{X \ar@{-}[d] \\ p}}
 \text{ if } b= \mbox{\textvisiblespace}
\]
where $*=\top$ if $p=q_f$ and $*=\bot$ otherwise. Note that these rules can only be applied to the
right branch of a configuration tree. For the left branch we add for every $a,c \in \Gamma$ and
$* \in \{\bot,\top\}$ the rules 
\[a \xrightarrow{(-,\bar a,*)} \epsilon \text{ and } X \xrightarrow{(-,\bar a,*)} X 
  \text{ if } a=\mbox{\textvisiblespace}, \]
as well as 
\[a \xrightarrow{(+,\bar c,*)}
  \vcenter{\xymatrix@R=0.3pc@C=0.2pc{a \ar@{-}[d] \\ c}} \]
and
\[ X  \xrightarrow{(+,\bar c,*)}  \vcenter{\xymatrix@R=0.3pc@C=0.2pc{X \ar@{-}[d] \\ c}} \text{ if } c \neq \mbox{\textvisiblespace} \text{ and } 
 X  \xrightarrow{(+,\bar c,*)} X \text{ if } c=\mbox{\textvisiblespace}.
\]
By construction, a path through the graph $G$ generated by $\mathcal R$ corresponds to a valid computation of $M$
started on the empty tape iff every transition with label $(+,a,*)$ respectively $(-,a,*)$ is followed by its 
counterpart labeled  $(+,\bar a,*)$ respectively $(-,\bar a, *)$. Let $H$ be the star graph with 
$|\Sigma|+1$ many vertices where the center vertex $v$ has  for every $(\$,a,*) \in \{+,-\} \times 
\Gamma \times\{\bot, \top\}$ a single outgoing edge with this label to a vertex $w$ and the single 
corresponding incoming edge from $w$ labeled $(\$,\bar a,*)$. 
If we define the synchronization constraint $C:=\{(\sigma, \sigma) \mid \sigma \in \Sigma\}$, the
synchronized product of $G$ and $H$ will contain exactly the valid computations of $M$. To decide
whether $M$ halts on the empty tape we thus have to check the truth of the formula 
\[
\exists xy \Big[\forall z \bigwedge_{\sigma \in \Sigma} \neg E_\sigma zx 
\wedge \exists z \Big(\Reach_\Sigma(x,z) \wedge \bigwedge_{\sigma \in \{+,-\} \times \bar \Gamma 
\times\{\top\}} E_\sigma zy\Big)\Big]
\]
in the semifinitely synchronized product of $G$ and $H$.
\qed

\section{Transitive Closure Logic over the Infinite Grid}

The infinite grid is the structure $\mathcal G = (\omega^2, S_1, S_2)$
with two successor relations $S_1$ and $S_2$. It can be viewed 
as the asynchronous and hence finitely synchronized product of two copies
of the natural numbers with successor relation, $\mathcal N_1=(\omega, S_1)$ and 
$\mathcal N_2=(\omega,S_2)$, defined by the empty synchronization constraint.

We show in this section how to interpret the first-order theory of 
addition and multiplication of the natural numbers  
in FO(TC)$_{(1)}^2$ (without parameters) over the infinite grid. 
FO(TC)$_{(1)}^2$ allows only transitive closure operators of arity one and 
a nesting depth of two. 

It is well known that the FO-theory of addition and multiplication 
of $\mathcal N$ is undecidable. However, since FO(TC)$_{(1)}$ can be interpreted
in MSO,  FO(TC)$_{(1)}$ is decidable over $\mathcal N$. From these results we can 
conclude that the  FO(TC)$_{(1)}^2$-theory is not preserved by finitely
synchronized products and thus obtain that we cannot extend FO(R)
much without losing decidability for finitely synchronized products.

To interpret the theory of addition and multiplication in FO(TC)$_{(1)}^2$
over the infinite grid we first connect the transitive closure theories of
$\mathcal N$ and $\mathcal G$.

\begin{lem}\label{lem:grid<->N} Let $k \ge 1$.
\begin{enumerate}[(a)] 
\item For every $\textrm{FO(TC)}_{(k)}^n$-sentence $\phi$ there is a 
$\textrm{FO(TC)}_{(2k)}^n$-sentence $\tilde \phi$ such that 
$\mathcal G \models \phi \Leftrightarrow \mathcal N \models \tilde \phi$.
\item 
For every $\textrm{FO(TC)}_{(2k)}^n$-sentence $\phi$ there is a 
$\textrm{FO(TC)}_{(k)}^n$-sentence $\hat \phi$ such that 
$\mathcal N \models \phi \Leftrightarrow \mathcal G \models \hat \phi$.
\end{enumerate}
\end{lem}
\proof
For (a) there is almost nothing to show. It suffices to split every variable
$x$ (interpreted as vertex of the grid) into coordinate variables $x_1$ and $x_2$
(interpreted as natural numbers) and to replace the atomic formulas
$S_1 xy$ by $S x_1 y_1$ and $S_2 x y$ by  $S x_2 y_2$. 

For (b) we identify every $x \in \omega$ with $(x,0) \in \omega^2$.
To reduce the number of variables needed in a TC operator 
we represent a pair of variables $x_1,x_2$ by a single variable 
$x=(x_1,x_2)$ to be interpreted as a vertex of the grid. 

To finish the proof it suffices to show that the following operations are
 $\textrm{FO(TC)}_{(1)}$ definable:
\begin{enumerate}[(i)]
\item $\pi_i$ with $\pi_1((x_1,x_2)):=(x_1,0)$ and $\pi_2((x_1,x_2)):=(0,x_2)$,
\item $\textrm{swap}_i$ with $\textrm{swap}_1((x,0)):=(0,x)$ and $\textrm{swap}_2((0,x)):=(x,0)$,
\item $\textrm{comb}$ with $\textrm{comb}((x,0),(0,y)):=(x,y)$
\end{enumerate}

Then a $\textrm{FO(TC)}_{(2k)}$ formula  
$[\TC_{\bar x, \bar y} \phi(\bar x,\bar y, \bar z)]\bar x, \bar y$
is equivalent to the $\textrm{FO(TC)}_{(k)}$ formula

\begin{multline*}
    \exists \bar u \bar v  \Big( \bigwedge_{1 \le i \le k}  \big(  u_i= \textrm{comb}(x_{2i-1},\textrm{swap}_2(x_{2i})) 
                         \wedge   v_i= \textrm{comb}(y_{2i-1},\textrm{swap}_2(y_{2i}))\big) \\
   \wedge   [\TC_{\bar u, \bar v} \tilde \phi(\bar u,\bar v, \bar z)]\bar u, \bar v \Big)
\end{multline*}
where
\begin{multline*}
  \tilde \phi := \exists \bar x \bar y  \Big( \bigwedge_{1 \le i \le k}  \big( x_{2i-1}= \pi_1(u_i) \wedge 
  x_{2i}=\textrm{swap}_2(\pi_2(u_i))  \wedge  y_{2i-1}= \pi_1(v_i) \\ 
  \wedge y_{2i}=\textrm{swap}_2(\pi_2(v_i)) \big) \wedge   \phi(\bar x, \bar y) \Big)
\end{multline*}
and in $\phi$ every occurrence of the symbol $S$ is replaced by $S_1$.

Let us now define the operations above:

\begin{eqnarray*}
\pi_1(x)=y & \leftrightarrow & y \le_2 x \wedge \forall z (z \le_2 x \rightarrow z=y) \\
\textrm{swap}_1(x)=y &\leftrightarrow &  \forall z (z \le_2 x \rightarrow z=x) \wedge  
  [\TC_{x,y} \exists z (S_1xz \wedge S_2yz)]x,y \\
 && \wedge \forall z (z \le_1 y \rightarrow z=y) \\
 \textrm{comb}(x,y)=z & \leftrightarrow & \forall u (u \le_2 x \rightarrow u=x) 
   \wedge \forall u (u \le_1 y \rightarrow u=y) \\
 && \wedge x \le_1 z \wedge y \le_2 z
\end{eqnarray*}
Observe that if the formula $\phi$ has no TC operators with parameters, then neither
$\tilde \phi$ nor $\hat \phi$ has (in $\tilde\phi$ only TC-formulas without parameters are introduced),
and that the nesting depth is not increased.
\qed

Let us now turn to the undecidability proof.

\begin{thm}
The $\textrm{FO(TC)}_{(1)}^2$-theory of the infinite grid is undecidable.
\end{thm}
\proof 
We define addition and multiplication in $\textrm{FO(TC)}_{(1)}$ over $\mathcal G$
without the use of parameters. By Lemma \ref{lem:grid<->N} it 
is enough to  define these operations in $\textrm{FO(TC)}_{(2)}$ over 
$\mathcal N$. The definition of addition is straightforward.
\[a+b=c  \leftrightarrow  \mathcal N \models [\TC_{x_1x_2,y_1 y_2} S x_1 y_1 \wedge S x_2 y_2] 0\,a, b\,c \]

To define multiplication note that $x \cdot y = \frac{(x+y)^2-x^2-y^2}{2}$,
hence it suffices to define the square function.  To define $x^2$ note that 
$x^2 = \sum_{i=0}^{x-1} 2i+1$. The formula
\[
\psi(x,y) =  [\TC_{x_1x_2,y_1 y_2} y_2=x_2+(x_2-x_1)+2 \\ 
 \wedge y_1=x_2]0\,1,xy
\]
defines all pairs of square numbers 
\[\big(\sum_{i=1}^{k-2} 2i+1, \sum_{i=1}^{k-1} 2i+1 \big) \text{ for } k \ge 3.\]
Hence $\mathcal N \models \psi[a,b]$ iff $b-a=2k-1$ for some $k \ge 2$.
Let 
\[\chi(x,y)= \exists z_1 \Big( \psi(z_1,y) \wedge \frac{y-z_1+1}{2}=x \Big).\]
Then $\mathcal N \models \chi[a,b]$ iff $b=a^2$.
\qed

A similar technique was used in \cite{av03} to define multiplication in $(\omega, +,0)$
using a transitive closure operator of arity one.

The nesting of transitive closure operators in the previous proof is necessary. If
we disallow nesting, even in the presence of parameters in the transitive closure 
formulas, the theory of the infinite grid is decidable.

\begin{thm}
The  $\textrm{FO(TC)}_{(1)}^1$-theory of the infinite grid is decidable.
\end{thm}
\proof
We reduce the $\textrm{FO(TC)}_{(1)}^1$-theory of the infinite grid $\mathcal G$ 
to Presburger arithmetic, the first-order theory of $\mathcal N_+ = (\omega, +, 0)$, in
the following sense: For every $\textrm{FO(TC)}_{(1)}^1$-formula $\phi(x_1,\ldots,x_n)$
one can construct  a Presburger formula $\tilde\phi(x_{11},x_{12},\ldots,x_{n1},x_{n2})$ such that
\begin{equation}\label{eq:fotc-grid-dec-1}
\mathcal G \models \phi[(k_1,l_1),\ldots,(k_n,l_n)] \Leftrightarrow \mathcal N_+ \models
\tilde \phi[k_1,l_1,\ldots,k_n,l_n].
\end{equation}
In order to construct $\tilde\phi$ it suffices to consider the case 
\[\phi(x_1,\ldots,x_n) =[\TC_{x_1,x_2} \psi(x_1,\ldots,x_n)]x_1,x_2,\] 
or for better readability 
\[\phi(x_1,\ldots,x_n) =[\TC_{x,y} \psi(x,y,x_3,\ldots,x_n)]x_1,x_2\] where $\psi$ is a 
first-order formula. The second notation emphasizes that $x_3,\ldots,x_n$ serve as 
parameters in the transitive closure formula.

In a first step we rewrite $\psi$ in a normal form, applying Hanf's Theorem for
first-order logic over graphs (see \cite{ha65,ef95,tho97a}).

For this purpose we recall some definitions. The $r$-sphere $r\sph(d)$
around a vertex $d \in \omega^2$ is the set of grid vertices which are of distance 
less or equal to $r$ from $d$, where we allow to traverse the edges in
either direction. Invoking the distributive normal form and Hanf's Theorem,
there exists a suitable $r >0$ such that $\psi(x_1,\ldots,x_n)$ is equivalent 
to a disjunction of formulas $\phi_\tau(x_1,\ldots,x_n)$ where each 
$\phi_\tau$ describes the isomorphism type $\tau$ of $\bigcup_{1 \le i \le n}
r\sph(c_i)$ for some tuple $c_1,\ldots,c_n$ of grid vertices. 
Let $T$ be the set of all such types. Since $T$ is finite
it suffices to consider only finitely many  tuples $c_1,\ldots,c_n$.

\emph{Remark.} In the general case, over an arbitrary graph instead of the 
infinite grid, Hanf's Theorem involves a statement on the number (up to a certain 
threshold) of spheres outside $\bigcup_{1 \le i \le n} r\sph(c_i)$. This 
statement is superfluous here due to the regular structure of the infinite grid.
(For technical convenience we assume that $(0,0)$ is included in the set of 
parameters, so every isomorphism type realizable in $\mathcal G$ outside 
$\bigcup_{1 \le i \le n} r\sph(c_i)$ occurs an infinite number of times.)

Due to the special structure of the grid, which we depict as a diagram 
with the bottom row and left column as margins, open upwards and 
to the right, every formula 
$\phi_\tau(x_1,\ldots, x_n)$ can be expressed by conditions on the vertices
$x_1,\ldots,x_n$ which fix their distances up to the radius $r$ from the
left margin as well as the bottom margin, and their relative distances up to $2r$.

It is convenient to express $\phi_\tau(x_1,\ldots,x_n)$ in terms of the $2n$
components of the vertices, obtaining a formula 
$\tilde \phi_\tau(x_{11},x_{12},\ldots,x_{n1},x_{n2})$
The formula $\tilde \phi_\tau$ is interpreted over $\omega$ and equivalent to $\phi$ in 
the sense of (\ref{eq:fotc-grid-dec-1}) above. It is a conjunction of statements

\begin{itemize}
\item $x_{ih}=k$ for $ 0 \le k \le r$ or $x_{ih} >r$
\item $(x_{i1},x_{i2}) = (x_{j1},x_{j2}) + (k,l)$ for $-2r \le k,l \le 2r$
\item $\dist((x_{i1},x_{i2}),(x_{j1},x_{j2})) >2r$
\end{itemize}
where $1 \le i,j \le n$ and $h \in \{1,2\}$. 

We now have to evaluate formulas of the form 
\begin{equation}\label{eq:fotc-grid-dec-2}
\Big[ \TC_{(x_{11},x_{12}),(x_{21},x_{22})} \\ \bigvee_{\tau \in T'} 
\tilde\phi_\tau(x_{11},x_{12},\ldots,x_{n1},x_{n2})\Big](s,t),(u,v)
\end{equation}
for some $T' \subseteq T$.

In a first step we note that it is possible to add disjuncts to (\ref{eq:fotc-grid-dec-2})
such that vertices tied to occur in a $2r$-sphere around a parameter $(x_{i1},x_{i2})$ for $i>2$
only need to appear as start vertex or as end vertex of any path described by (\ref{eq:fotc-grid-dec-2}).
Hence vertices tied to parameters can be handled without the use of TC, by an appropriate 
modification of the formula.

Let $I$ be an initial segment of the grid encompassing the $2r$-spheres around parameters 
$(x_{i1},x_{i2})$ for $i>2$. Outside this initial segment, in a second step, it suffices to 
consider formulas (\ref{eq:fotc-grid-dec-2}) in which only type formulas $\tilde\phi_\tau$
which contain

\[
 x_{11} = k_1  \wedge x_{12}>r  \text{ or }  x_{11}>r  \wedge x_{12}=k_2 \text{ or }  x_{11} >r  \wedge x_{12}>r 
 \text{ for } k_1,k_2 \le r
\]
and 
\[
 x_{21}=l_1 \wedge x_{22}>r  \text{ or }  x_{21}>r  \wedge x_{22}=l_2 \\
 \text{ or }  x_{21} >r  \wedge x_{22}>r 
 \text{ for } l_1,l_2 \le r
\] 
and
\[
 \dist((x_{11},x_{12}),(x_{21},x_{22})) >2r 
 \text{ or } (x_{11},x_{12}) = (x_{21},x_{22})+(k,l) 
 \text{ for } -2r \le k,l \le 2r
\] 
appear.

It is now possible to apply a finite saturation process to obtain a formula
\begin{equation}\label{eq:fotc-grid-dec-3}
\Big[ \TC_{(x_{11},x_{12}),(x_{21},x_{22})} \bigvee_{1 \le j \le m} 
\tilde\phi_j(x_{11},x_{12},\ldots,x_{n1},x_{n2})\Big](s,t),(u,v)
\end{equation}
which is equivalent to (\ref{eq:fotc-grid-dec-2}) and where TC and $\bigvee$ commute, i.e.
\begin{multline*}
 \mathcal G \models \Big[\TC_{(x_{11},x_{12}),(x_{21},x_{22})} \bigvee_{1 \le j \le m} \tilde\phi_j \Big](s,t),(u,v)
   \Leftrightarrow \\
 \mathcal G \models \bigvee_{1 \le j \le m} \Big[ \TC_{(x_{11},x_{12}),(x_{21},x_{22})} \tilde\phi_j \Big](s,t),(u,v).
\end{multline*}
 
The subformulas $\tilde\phi_j$ in (\ref{eq:fotc-grid-dec-3}) have the same format as  
the subformulas $\tilde\phi_\tau$ in (\ref{eq:fotc-grid-dec-2}) except that the center of 
the excluded $2r$-sphere around $(x_{11},x_{12})$ may be shifted by a bounded distance 
from $(x_{11},x_{12})$ or be missing, or $\tilde\phi_\tau$ defines the complete relation 
outside $I$ and the border stripes of width $r$.
Thus it remains to consider two cases.

\noindent \emph{Case 1.}  If $\tilde\phi_j$ contains a conjunct 
excluding some $2r$-sphere then the relation defined by $[\TC_{(x_{11},x_{12}),(x_{21},x_{22})} \tilde\phi_j](s,t),(u,v)$ 
is cofinite (w.r.t. the grid excluding $I$ and border stripes of width $r$, or a fixed line in one of the border stripes) 
and hence definable without the use of a transitive closure operator.

\noindent\emph{Case 2.} If  $\tilde\phi_j$ fixes relations of the form
\begin{equation}\label{eq:fotc-grid-dec-4}
 (x_{21},x_{22})=(x_{11},x_{12})+(k_i,l_i) 
\end{equation}
for $i=1,\ldots,N$  and $-2r\le k_i,l_i \le 2r$.
the formula  \[[\TC_{(x_{11},x_{12}),(x_{21},x_{22})} \tilde\phi_j](s,t),(u,v)\] 
expresses that there is a path from $(s,t)$ to $(u,v)$ consisting of steps of the form 
(\ref{eq:fotc-grid-dec-3}). The set of vertices $(u,v)$ reachable in this way from $(s,t)$ 
can be represented as the union of paths in the finite initial segment $I$ of the grid 
and finitely many sets of the form \[\{(u,v)\mid (u,v)=(s',t')+y_1 (k_1,l_1)+ \ldots + y_N (k_N,l_N)\}.\]
Here $y_i \ge 0$, the $(s',t')$ range over boundary vertices of $I$, 
and the $(k_i,l_i)$ are from  (\ref{eq:fotc-grid-dec-4}). It follows that 
the relation defined by (\ref{eq:fotc-grid-dec-2}) is definable in Presburger arithmetic.
\qed

\section{Conclusion}







We have proved a result on compositional model checking for 
a logic including reachability predicates, and we have shown 
tight limitations for possible extensions of this result.

Let us mention some questions left open in this paper: 

\begin{enumerate}
\item The composition result (Theorem 3.1) 
should be generalized to infinite products. 
\item For an extension of Theorem 3.1, one can enrich FO(R) by an operator for 
``recurrent reachability'' (existence of an infinite path 
which visits a designated set infinitely often), or one 
can consider stronger logics like (fragments of) CTL.
\item Interesting subcases of Theorem 3.1 should be found where 
the mentioned blow-up of complexity can be avoided.
\item The distinction between products which are asynchronous, 
finitely synchronized, or synchronized should be refined, 
by allowing other means of coordination between component 
structures, also incorporating the special case of synchronization 
of parameterized systems composed from identical components.
\end{enumerate}

\section*{Acknowledgment}

We thank C. L\"oding for pointing us to GTRS-graphs to prove 
Theorem \ref{theo:semi-undec} and
the anonymous referees (both of the conference version and the 
journal version of this paper) for many helpful comments and 
pointers to related literature.


\bibliographystyle{alpha}

\bibliography{literatur}

\end{document}